\documentclass[reprint, amsmath, amssymb, aps, prd, longbibliography]{revtex4-2}

\usepackage{graphicx}
\usepackage{dcolumn}
\usepackage{bm}
\usepackage{hyperref}
\usepackage{scrhack}
\usepackage{float}
\usepackage{makecell}
\usepackage{booktabs}
\usepackage{boldline}

\begin{document}

\title{Earth-Density Effects in Long Baseline Neutrino Experiments}

\author{Tia Pandit}
\email{tiapandit777@outlook.com}
\affiliation{Department of Physics, Kirori Mal College, University of Delhi, Delhi-110007, India}

\author{Bipin Singh Koranga}
\email{bskoranga@kmc.du.ac.in}
\affiliation{Department of Physics, Kirori Mal College, University of Delhi, Delhi-110007, India}

\date{\today}
\begin{abstract}
The accurate modelling of Earth matter effects is a critical systematic consideration for precision measurements of the CP-violating phase $\delta_{CP}$ in long-baseline (LBL) neutrino oscillation experiments. We investigate the extent to which constant and path-averaged density approximations are sufficient for reproducing the $\nu_\mu \rightarrow \nu_e$ appearance probabilities predicted by the Preliminary Reference Earth Model (PREM) across baselines ranging from $L = 1000$ km to $L = 12000$ km. Using a full three-flavour matrix exponentiation framework with spatially resolved PREM density profiles, we demonstrate that the 
constant density approximation introduces a negligible bias of less than $0.3^{\circ}$ in the reconstructed $\delta_{CP}$ for baselines up to $L \approx 5000$ km, but undergoes a sharp increase beyond this threshold, reaching $17.8^{\circ}$ at $L = 7000$ km and $172.2^{\circ}$ at $L = 12000$ km. We show that this bias arises from energy-dependent distortions in the oscillation probability produced by the Earth's internal density layering, which generate degeneracies between matter-induced and intrinsic CP-violating contributions to the appearance channel that cannot be removed by marginalising over a single effective density. These results demonstrate that the constant density approximation is not a conservative simplification but a source of fundamental systematic error at long baselines, and motivate the incorporation of spatially resolved Earth density treatments, such as PREM, in the analysis frameworks of next-generation long-baseline neutrino oscillation experiments.
\end{abstract}

\maketitle

\section{Introduction}
Neutrinos are electrically neutral, spin 1/2 fermions that interact only via weak nuclear force \& gravity. They're produced as a result of nuclear processes in the Sun, supernovae, reactors and accelerators. Neutrinos exist in three flavours;  electron neutrino $(\nu_e)$, muon neutrino $(\nu_\mu)$ and tau neutrino $(\nu_\tau)$ which undergo flavour oscillations \cite{AkhmedovNiro} during propagation, implying non-zero mass and physics beyond the Standard Model.\\
When neutrinos propagate through matter such as the Earth's crust, electrons in matter interact only with electron neutrinos ($\nu_e$) via charged-current processes, while muon and tau neutrinos ($\nu_\mu$, $\nu_\tau$) do not experience this interaction channel. This coherent, forward scattering of $\nu_e$ creates an additional effective potential that shifts the relative energies, effective masses, and mixing angles of the neutrino states. Since oscillations depend on these quantities, even a modest matter-induced modification can substantially alter the flavour evolution. This phenomenon is known as the\textbf{ Mikheyev--Smirnov--Wolfenstein (MSW) matter effect} \cite{Wolfenstein1978,SmirnovMSW}.
\\ 
\\
In LBL neutrino experiments \cite{Diwan2016} such as T2K \cite{Abe2011T2K}, NO$\nu$A \cite{Bian_2013_NOvA}, and DUNE \cite{DUNE_CDR_Physics_2016}, neutrinos travel hundreds to thousands of kilometres through Earth. Over such distances, the MSW effect has enough baseline to produce observable changes in oscillation probabilities \cite{koranga2009determiningsigndelta31future}. For an initial $\nu_\mu$ beam, the matter environment alters both the amplitude and phase of its oscillation into $\nu_e$ by the time it reaches the detector \cite{NOvA, Raut2017}. CP-violating effects in neutrino oscillations are fundamental, coming from the laws of particle physics themselves \cite{Dick1999CP, koranga2013}. If CP violation is present, neutrinos and antineutrinos behave differently even in empty space. This difference is intrinsic and does not depend on where the neutrinos travel.
In contrast, matter-induced effects are environmental. They arise because neutrinos pass through matter, like the Earth, which contains electrons but not positrons. This uneven background enhances the $\nu_\mu \rightarrow \nu_e$ appearance probability through the MSW effect, while for antineutrinos the effective matter potential changes sign, leading to a suppression of the $\bar{\nu}_\mu \rightarrow \bar{\nu}_e$ transition. This neutrino–antineutrino asymmetry is hence a consequence of matter-induced effects rather than a direct signal of CP violation.
The matter-induced asymmetry has a predictable energy and baseline dependence that LBL experiments capitalize on. They differentiate it from the symmetry breaking caused by intrinsic CP violation. This is done by measuring oscillation probabilities across a range of energies and running in both neutrino and antineutrino modes. In this way, the Earth itself becomes a useful probe, allowing long-baseline experiments to determine the neutrino mass ordering while simultaneously constraining the CP-violating phase.

\subsection{Three-Flavour Formulation}
The three flavour states are quantum superpositions of mass eigenstates and the relation between them is given by the Pontecorvo--Maki--Nakagawa--Sakata (PMNS) matrix $U$ \cite{OhlssonSnellman2000},
\begin{equation}
|\nu_\alpha\rangle = \sum_{i=1}^{3} U_{\alpha i}\,|\nu_i\rangle,
\qquad \alpha = e,\mu,\tau .
\end{equation}
This mismatch between flavor and mass eigenstates leads to neutrino flavor oscillations during propagation. The oscillation phenomenon depends on three mixing angles $(\theta_{12},\theta_{23},\theta_{13})$, a CP-violating phase $\delta_{\rm CP}$, and the mass-squared differences $\Delta m_{21}^2$ and $\Delta m_{31}^2$, and its observation provides direct evidence for physics beyond the Standard Model.
\begin{widetext}
\[
U =
\begin{pmatrix}
c_{12} c_{13} &
s_{12} c_{13} &
s_{13} e^{-i \delta_{\rm CP}} \\[0.4em]

- s_{12} c_{23} - c_{12} s_{23} s_{13} e^{i \delta_{\rm CP}} &
\;\; c_{12} c_{23} - s_{12} s_{23} s_{13} e^{i \delta_{\rm CP}} &
s_{23} c_{13} \\[0.4em]

\;\; s_{12} s_{23} - c_{12} c_{23} s_{13} e^{i \delta_{\rm CP}} &
- c_{12} s_{23} - s_{12} c_{23} s_{13} e^{i \delta_{\rm CP}} &
c_{23} c_{13}
\end{pmatrix}
\]
where $c_{ij} \equiv \cos\theta_{ij}$ and $s_{ij} \equiv \sin\theta_{ij}$.
\\
\end{widetext}
Neutrino oscillations are described as the quantum mechanical time evolution of a three-component state vector. In natural units $( \hbar = c = 1)$, it is governed by a Schr\"odinger-like equation,
\begin{equation}
i\,\frac{d}{dt}\,|\nu(t)\rangle = H\,|\nu(t)\rangle ,
\end{equation}
where $H$ is the effective Hamiltonian. In the relativistic limit, the propagation time $t$ is identified with the baseline $L$.
In vacuum, the Hamiltonian is diagonal in the mass basis,
\begin{equation}
H_{\rm vac}^{(m)} = \begin{pmatrix}
E_1 & 0 & 0 \\
0 & E_2 & 0 \\
0 & 0 & E_3
\end{pmatrix}
\end{equation}
with $E_a \simeq E + m_a^2/(2E)$, while in the flavor basis it is given by
\begin{equation}
   H_{\rm vac}^{(f)} = U H_{\rm vac}^{(m)} U^{-1} 
\end{equation}
When neutrinos propagate through matter, the MSW effect induces an additional potential~\cite{OhlssonSnellman2000} that is diagonal in the flavor basis,
\begin{equation}
V_f = \begin{pmatrix}
A & 0 & 0 \\
0 & 0 & 0 \\
0 & 0 & 0
\end{pmatrix}, \qquad A = \sqrt{2}\,G_F N_e ,
\end{equation}
where $N_e$ denotes the electron number density.
The total Hamiltonian in matter is therefore
\begin{equation}
H_f = H_{\rm vac}^{(f)} + V_f ,
\end{equation}
or, equivalently, in the mass basis,
\begin{equation}
H_m = H_{\rm vac}^{(m)} + U^{-1} V_f U .
\end{equation}
In the three-flavor case, the Hamiltonian is non-diagonal in both bases, and the oscillation
probabilities are obtained by diagonalizing $H_m$ and constructing the corresponding
evolution operator. Matter effects thus modify both the effective mass-squared differences
and mixing angles, leading to resonant enhancement of oscillations.

\section{Role of Earth Density Uncertainties}
The electron number density is determined by the Earth matter profile \cite{giunti2007neutrino}:
\begin{equation}
   N_e(x) = Y_e(x)\,\frac{\rho(x)}{m_p}
\end{equation}
where $\rho(x)$ is the Earth density along the neutrino trajectory, 
$Y_e(x)$ is the electron fraction (having a standard value of 0.5 for terrestrial matter) , and $m_p$ is the proton mass.
It is clear that the electron number density is directly proportional to Earth density, rendering the MSW potential $(A = \sqrt{2}\,G_F N_e)$ position-dependent.
\\
The Earth is radially stratified, in essence density increases with depth and neutrinos crossing hundreds to thousands of kilometers sample multiple layers, not a uniform medium. For long baselines, neutrinos mostly travel through the mantle, not the core but even within the mantle the density is not constant. Constant density approximations make use of the fact that Earth density varies slowly along the path, keeping the propagation mostly adiabatic but as LBL experiments enter precision era, “mostly” is not “exactly”. Earth density is no longer a background detail but a systematic uncertainty that competes with CP sensitivity \cite{Kelly_Parke_2018_DUNE_Matter_Profile}.
Even modest mismodelling of the matter profile can introduce biases that imitate or obscure genuine CP-violating effects.
\\
\\
Developed by Dziewonski \& Anderson in 1981, Preliminary reference Earth model \cite{dziewonski1981} is the standard model of Earth's internal structure used across geophysics and, increasingly, neutrino physics. PREM gives us density, seismic velocities, and pressure as a function of depth from the surface to the core. The Earth is divided into shells with varying densities across major internal layers, a simplified version of which is described in Table I.
\begin{table}[h]
\centering
\caption{Density profile of the Earth based on the Preliminary Reference Earth Model (PREM), showing variations across major internal layers.}
\begin{tabular}{lcc}
\hline
\textbf{Region} & \textbf{Depth (km)} & \textbf{Density (g/cm$^3$)} \\
\hline
Crust        & 0--35         & $\sim$2.9   \\
Upper Mantle & 35--660       & 3.3--3.9    \\
Lower Mantle & 660--2891     & 4.4--5.6    \\
Outer Core   & 2891--5150    & 9.9--12.2   \\
Inner Core   & 5150--6371    & $\sim$13    \\
\hline
\end{tabular}
\label{tab:prem}
\end{table}
\\
As established previously, the strength of MSW Effect depends directly on electron density, which tracks matter density from PREM. At long baselines like 5000–12000 km, neutrinos pass through multiple density shells and hence, single average density is a generally poor approximation for such baselines.
\section{Quantitative Analysis}
To quantify the impact of the Earth density profile on $\delta_{CP}$ 
reconstruction, we simulate $\nu_\mu \rightarrow \nu_e$ appearance 
event rates across a range of baselines and compare the results 
obtained under the full PREM profile against those from the constant 
density approximation.\\
\begin{figure}[h!]
  \centering
    \includegraphics[width=1\linewidth]{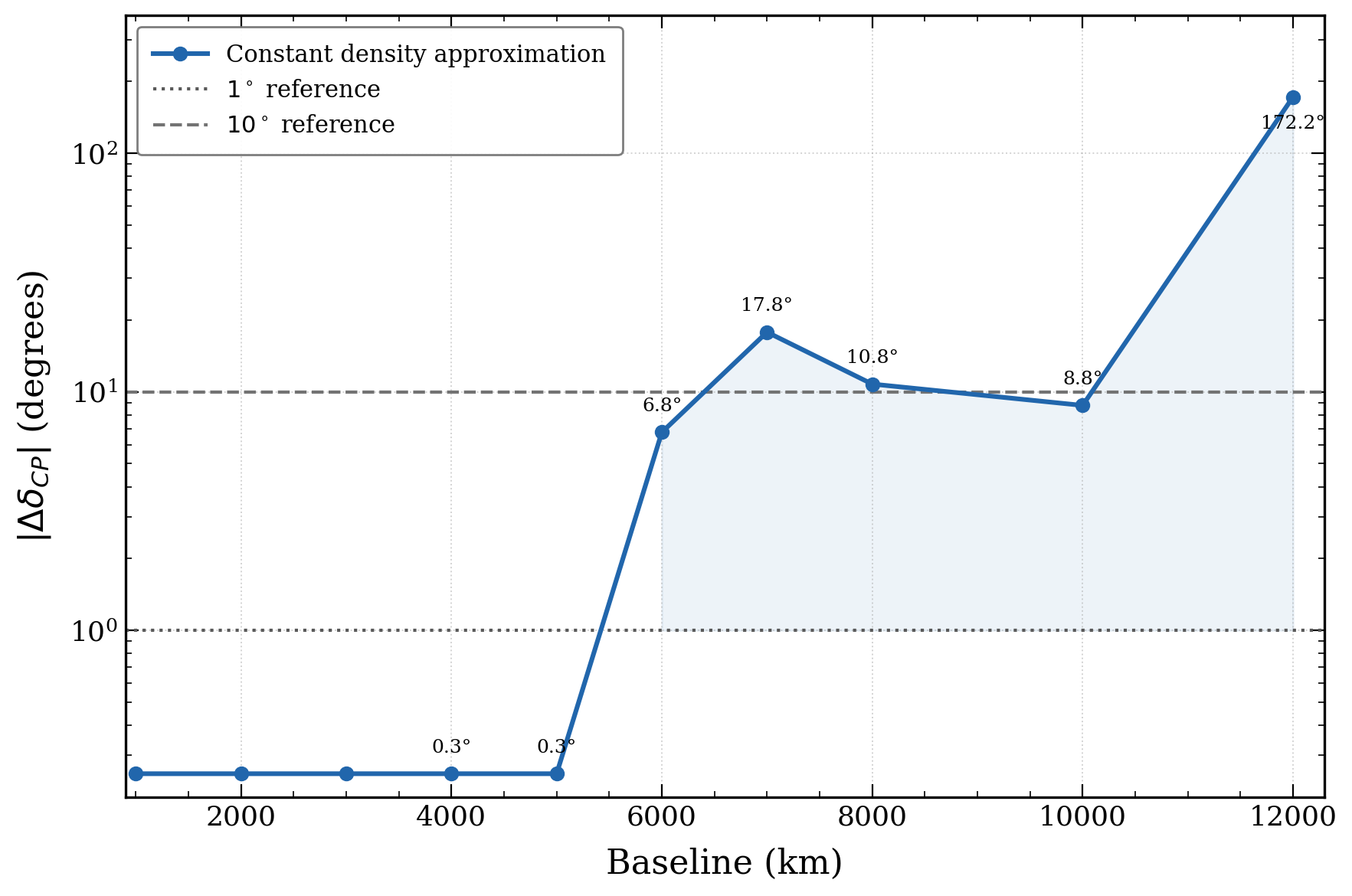}
    \caption{Absolute bias in the reconstructed CP phase $|\Delta\delta_{CP}|$ introduced by the constant density approximation, shown as a function of baseline $L$. The true event rates are generated using the full PREM Earth density profile at $\delta_{CP} = -90^{\circ}$, while the constant density model uses the path-length-weighted average density along each trajectory. The bias remains below $1^{\circ}$ for baselines up to $L \approx 5000$ km, but grows sharply beyond this threshold, reaching $17.8^{\circ}$ at $L = 7000$ km and $172.2^{\circ}$ at $L = 12000$ km, where the neutrino trajectory begins to sample the high-density outer core.}
        \label{fig:biasVSbaseline}
\end{figure}\\
The results of Figure \ref{fig:biasVSbaseline} reveal that the constant density approximation introduces a negligible bias of less than $0.3^{\circ}$ for baselines up to $L = 5000$ km, but undergoes a sharp increase beyond this threshold, reaching $17.8^{\circ}$ at $L = 7000$ km and $172.2^{\circ}$ at $L = 12000$ km. This threshold corresponds physically to the onset of sensitivity to the high-density lower mantle and, at the longest baselines, the outer core, where the electron density departs significantly from any single average value. Notably, the bias is non-monotonic, decreasing from $17.8^{\circ}$ to $8.8^{\circ}$ between $L = 7000$ km and $L = 9000$ km. This is 
attributed to oscillation phase degeneracies at certain baselines where 
the constant density $\chi^2$ minimum accidentally coincides with the 
true value. 
\begin{table}[H]
\centering
\caption{Path-averaged density, best-fit $\delta_{CP}$ under the 
constant density approximation, and resulting bias relative to the 
true value of $\delta_{CP} = -90^{\circ}$, as a function of baseline.}
\begin{tabular}{cccc}
\toprule
\textbf{Baseline (km)} & \textbf{$\rho_{\text{avg}}$ (g/cm$^3$)} & \textbf{Const. best-fit ($^{\circ}$)} & \textbf{Bias ($^{\circ}$)} \\
\midrule
1000  & 3.300 & $-89.7$  & 0.3   \\
2000  & 3.300 & $-89.7$  & 0.3   \\
3000  & 3.300 & $-89.7$  & 0.3   \\
4000  & 3.300 & $-89.7$  & 0.3   \\
5000  & 3.300 & $-89.7$  & 0.3   \\
7000  & 4.289 & $-107.8$ & 17.8  \\
9000  & 4.616 & $-98.8$  & 8.8   \\
12000 & 7.547 & $97.8$   & 172.2 \\
\bottomrule
\end{tabular}
\label{tab:bias}
\end{table}

The sharp growth of bias at long baselines demonstrates that the constant density approximation is not a conservative simplification but a source of fundamental systematic error, capable of reconstructing the wrong sign and magnitude of CP violation entirely.

\begin{figure}[h!]
  \centering
    \includegraphics[width=1\linewidth]{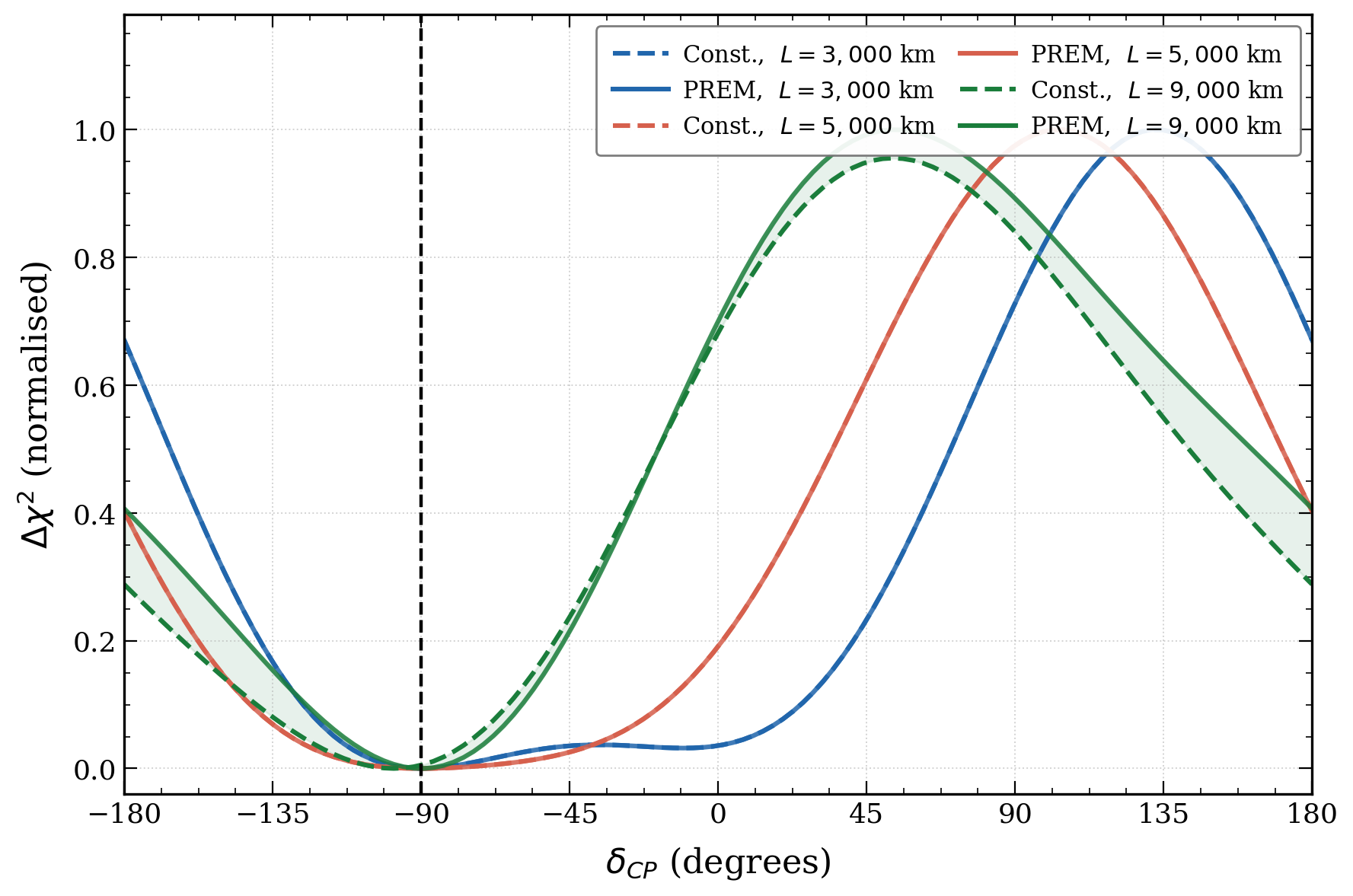}
    \caption{Normalised $\Delta\chi^2$ profiles as a function of the test CP phase $\delta_{CP}$ for baselines of $L=3000$ km, $L=5000$ km, and $L=9000$ km. In all cases, the true event spectra are generated using the PREM Earth density profile with $\delta_{CP}^{\rm true}=-90^\circ$. The solid curves correspond to PREM-based test spectra fitted against PREM-generated true spectra, while the dashed curves correspond to constant-density test spectra fitted against the same PREM-generated true spectra. Thus, the dashed curves quantify the bias introduced by replacing the realistic Earth density profile with a constant-density approximation in the fit. For each baseline, the curves are independently normalised to their respective maxima to facilitate direct comparison of profile shapes across different baselines. The shaded regions indicate the differences between the PREM and constant-density treatments, illustrating the increasing impact of Earth matter modeling at longer baselines. The vertical dashed line marks the true value, $\delta_{CP}=-90^\circ$.
}
    \label{fig:Chi_Squared}
\end{figure}
Figure \ref{fig:Chi_Squared} clarifies several key features of the constant density approximation's failure as a function of baseline. At $L = 3000$ km, the constant density and PREM profiles are nearly indistinguishable, with both curves minimising at $\delta_{CP} \approx -90^{\circ}$ and the shaded discrepancy region remaining negligibly small. This confirms that the constant density approximation is adequate for short-to-medium baseline experiments. At $L = 5000$ km, a modest but visible separation begins to emerge between the two profiles, particularly in the wings of the $\Delta\chi^2$ curve, though both minima remain close to the true 
value. The situation changes markedly at $L = 9000$ km, where the 
constant density minimum shifts visibly away from $-90^{\circ}$ and 
the shaded discrepancy region broadens substantially across the full 
range of $\delta_{CP}$. Critically, at this baseline the two profiles exhibit not only different minimum locations but different overall shapes, indicating that the constant density approximation is not merely introducing a shift in the reconstructed $\delta_{CP}$ but is misrepresenting the underlying $\chi^2$ landscape entirely. This shape distortion would affect not only the best-fit value of 
$\delta_{CP}$ but also the inferred confidence intervals, potentially leading to an incorrect assessment of the statistical significance of a CP violation measurement.\\
\begin{figure}[h!]
  \centering
    \includegraphics[width=1\linewidth]{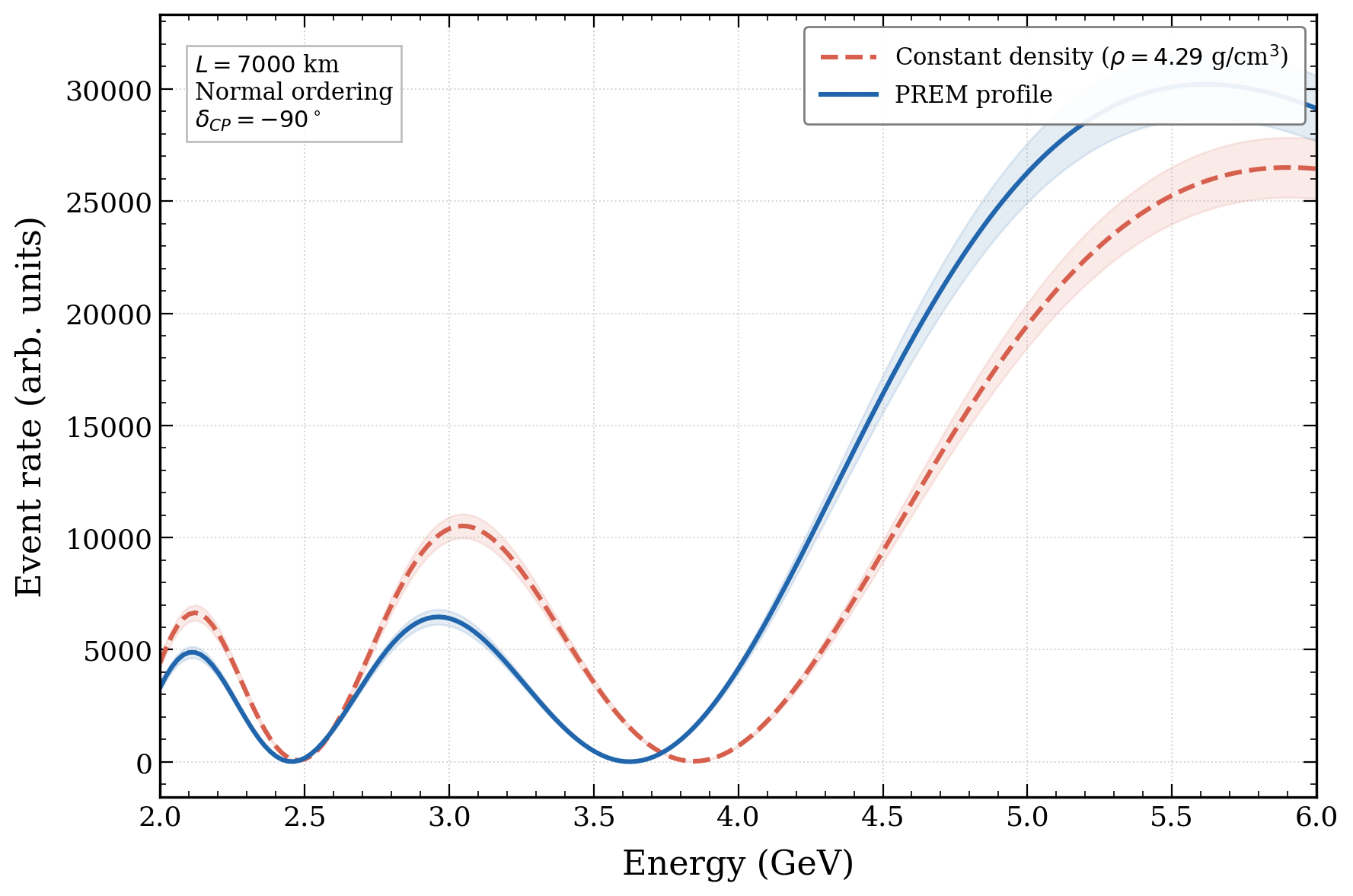}
    \caption{Predicted $\nu_\mu \rightarrow \nu_e$ appearance event rates as a function of reconstructed neutrino energy at a baseline of $L = 7000$ km, for $\delta_{CP} = -90^{\circ}$ under normal mass ordering. Results are shown for both the constant density approximation (dashed red) and the full PREM Earth density profile (solid blue). Shaded bands represent the total $1\sigma$ uncertainty, combining statistical uncertainty $(\sqrt{N})$ with systematic contributions from flux normalisation, interaction cross-section modelling, and an assumed $5\%$ overall systematic uncertainty.}
    \label{fig:EventRateVSEnergy}
\end{figure}
\\
In Figure \ref{fig:EventRateVSEnergy}, at $L = 7000$ km, the two models exhibit significant structural differences across the full energy range of $2$--$6$ GeV. The constant density approximation predicts a sharper and more pronounced primary oscillation peak near $E \approx 3.0$ GeV compared to the PREM profile, which predicts a broader and more suppressed peak at the same energy. Beyond $E \approx 3.7$ GeV however, the behaviour reverses; the PREM profile rises more steeply and reaches substantially higher event rates above $4$ GeV, reflecting the enhanced and spatially varying matter potential accumulated as the neutrino trajectory penetrates the denser lower mantle at this baseline. The systematic divergence between the two models across nearly the entire energy spectrum, well beyond the respective $1\sigma$ uncertainty bands, confirms that at $L = 7000$ km the constant density approximation is statistically distinguishable from the true Earth profile, and its use would introduce a significant and energy-dependent bias into any measurement of $\delta_{CP}$.

\subsection*{Why $7000$ km?}
In the previous plot, we have chosen a baseline length of $7000$ km as it becomes the first point where the constant density approximation demonstrably fails, making it the most scientifically interesting case to study in detail. At $7000$ km, the chord through the Earth (at trajectory midpoint, y = 0) reaches a minimum radius of:
\begin{equation}
    r_{\min} = \sqrt{R_\oplus^2 - (L/2)^2} = \sqrt{6371^2 - 3500^2} \approx 5300 \text{ km}
\end{equation}
This places the deepest point of the trajectory in the lower mantle at $\rho \approx 5 \text{ g/cm}^3$, significantly denser than the upper mantle. A single average density cannot capture this variation. Lastly, at $7000$ km the trajectory does not yet reach the outer core, meaning the dramatic bias at this baseline is caused purely by lower mantle density variation, a cleaner and more convincing demonstration than at 12000 km where core-crossing introduces additional complexity.
\subsection*{Why $\delta_{CP} = -90^{\circ}$?}
A true CP phase of $\delta_{CP} = -90^{\circ}$ is adopted as it 
maximises the leptonic CP asymmetry ($\sin\delta_{CP} = -1$) \cite{Kimura2002}, 
is consistent with current experimental hints from T2K and 
NO$\nu$A, and represents the regime in which mismodelling of 
the Earth matter potential produces the largest and most 
experimentally consequential bias in $\delta_{CP}$ reconstruction.

\subsection*{Methodology of Figures}

In Figure 1, to quantify the bias introduced by the constant density approximation as a function of baseline, we simulated $\nu_\mu \rightarrow \nu_e$ appearance event rates across eight baselines ranging from $L = 1000$ km to $L = 12000$ km, using the full PREM Earth 
density profile as the ground truth at $\delta_{CP} = -90^{\circ}$ 
under normal mass ordering. For each baseline, the path-length-weighted 
average density was computed from the PREM profile along the 
corresponding chord through the Earth, and this single value was used 
as the constant density approximation, representing the most 
physically motivated and optimistic version of the approximation. A 
Poisson $\chi^2$ statistic was minimised over $\delta_{CP} \in 
[-180^{\circ}, 180^{\circ}]$ using the constant density model, and the 
deviation of the resulting best-fit value from the true value of 
$-90^{\circ}$ was recorded as the absolute bias $|\Delta\delta_{CP}|$.
\\
\\
The $\Delta\chi^2$ profiles shown in Figure 2 are constructed by 
simulating $\nu_\mu \rightarrow \nu_e$ appearance event rates across 
a neutrino energy range of $2$--$6$ GeV for three baselines, $L = 
3000$ km, $L = 5000$ km, and $L = 9000$ km, under both the full PREM 
Earth density profile and the constant density approximation. For each 
baseline, the PREM profile is obtained by mapping the neutrino 
trajectory as a chord through the Earth using radial geometry, sampling 
the density at 300 equally spaced steps along the path and assigning 
the appropriate shell density from a four-layer Earth model comprising 
the inner core ($\rho = 13.0$ g/cm$^3$), outer core ($\rho = 11.0$ 
g/cm$^3$), lower mantle ($\rho = 5.0$ g/cm$^3$), and upper mantle 
and crust ($\rho = 3.3$ g/cm$^3$). The constant density approximation 
uses the path-length-weighted average of this profile along each 
trajectory, representing the most physically motivated single-density 
estimate. Neutrino propagation is handled by numerically solving the 
Schr\"{o}dinger-like evolution equation in the flavour basis, with the 
full three-flavour Hamiltonian comprising a vacuum term constructed 
from the PMNS matrix and mass-squared differences, and a matter 
potential term accounting for coherent forward scattering of electron 
neutrinos off ambient electrons via the MSW effect. At each step along 
the trajectory, the propagator $U = \exp(-i \, \mathcal{H} \, \Delta x)$ 
is computed via matrix exponentiation and applied to the flavour state 
vector, with the procedure repeated independently for neutrinos and 
antineutrinos. The appearance probability $P(\nu_\mu \rightarrow \nu_e)$ 
is then weighted by a model beam flux $\phi(E) \propto e^{-E/3}$, a 
linear neutrino interaction cross-section $\sigma(E) \propto E$, and 
a flat detector efficiency of $80\%$ to produce event rate spectra. 
A Pearson-type $\chi^2$ statistic with regularization of the form
\begin{equation}
    \Delta\chi^2 = \sum_i 
    \frac{\left(N_i^{\text{test}} - N_i^{\text{true}}\right)^2}
    {N_i^{\text{true}} + 1}
\end{equation}
is evaluated by comparing test event rates, computed over a grid of 
120 values of $\delta_{CP} \in [-180^{\circ}, 180^{\circ}]$, against 
the true rates generated at $\delta_{CP} = -90^{\circ}$ using the PREM 
profile. Both neutrino and antineutrino channels are included in the 
$\chi^2$ sum to break the $\delta_{CP}$--matter effect degeneracy. The 
resulting $\Delta\chi^2$ curves are independently normalised to their 
respective maxima to allow direct shape and minimum-location comparison 
across baselines and density models.
\\
\\
The predicted $\nu_\mu \rightarrow \nu_e$ appearance event rates shown 
in Figure 3 are computed at a fixed baseline of $L = 7000$ km across 
200 energy bins spanning $2$--$6$ GeV, under both the full PREM Earth 
density profile and a constant density approximation of $\bar{\rho} = 
4.289$ g/cm$^3$, corresponding to the path-length-weighted average 
density along the $L = 7000$ km trajectory. The neutrino trajectory is 
discretised into 400 equal steps, with the local PREM density at each 
step assigned via a four-shell Earth model using the radial coordinate 
computed from chord geometry. At each step, the three-flavour 
Hamiltonian is constructed from the PMNS matrix evaluated at 
$\delta_{CP} = -90^{\circ}$, the vacuum mass-squared differences, and 
the MSW matter potential $V = 7.56 \times 10^{-14}\,\rho\, Y_e$ eV, 
and the neutrino flavour state is evolved via matrix exponentiation 
from an initial pure $\nu_\mu$ state. The resulting appearance 
probability is weighted by a falling beam flux $\phi(E) \propto 
e^{-E/3}$, a linear cross-section $\sigma(E) \propto E$, and an 
exposure factor $\mathcal{N} = 10^5$ to produce event rates, with 
$1\sigma$ uncertainty bands combining statistical uncertainty 
$\sqrt{N}$ and a $5\%$ systematic uncertainty in quadrature.\\
The discussion so far clearly highlights the need for refined geophysical inputs and quantified density uncertainties in high-precision analyses. Understanding how matter-induced distortions affect appearance probabilities is essential to ensure that observed neutrino--antineutrino differences reflect fundamental physics rather than environmental assumptions.

\section{Discussion \& Future Directions}
This analysis demonstrates that treating Earth matter effects with a 
constant or path-averaged density is no longer sufficient for precision 
long-baseline neutrino experiments. Although such approximations capture 
the leading MSW enhancement, they miss the energy-dependent distortions 
produced by realistic spatial variations in the Earth's density, as 
evidenced by the systematic divergence between the constant density and 
PREM event rate spectra visible in Figure 3 at $L = 7000$ km.

A key consequence of matter-density mismodelling is the introduction of 
energy-dependent degeneracies between intrinsic CP-violating effects and 
matter-induced contributions in the $\nu_\mu \rightarrow \nu_e$ 
appearance channel. As shown in Figure 2, these degeneracies manifest 
as a growing separation between the $\Delta\chi^2$ minima of the 
constant density and PREM profiles with increasing baseline, negligible at $L = 3000$ km but clearly visible at $L = 9000$ km, confirming that they cannot be eliminated by marginalising over a single density normalisation. The result is a CP sensitivity that varies strongly with energy, where in some regimes the limiting factor is no longer detector performance or exposure, but uncertainty in the Earth density model itself.

Figure 1 quantifies this effect directly, revealing that the bias in 
the reconstructed $\delta_{CP}$ introduced by the constant density 
approximation remains below $1^{\circ}$ for baselines up to $L \approx 
5000$ km but grows sharply beyond this threshold, reaching $17.8^{\circ}$ 
at $L = 7000$ km and $172.2^{\circ}$ at $L = 12000$ km. The 
non-monotonic behaviour between $7000$ km and $9000$ km is attributed 
to oscillation phase degeneracies, and the catastrophic bias at $12000$ 
km, where the neutrino trajectory begins sampling the high-density 
outer core, highlights that at such baselines the constant density 
approximation does not merely shift the reconstructed $\delta_{CP}$ 
but reconstructs the wrong sign of CP violation entirely.

Hence, the primary goal for future long-baseline analyses is to shift 
the treatment of Earth's matter density from a simple numerical constant 
to a spatially-resolved systematic uncertainty. By doing so, experiments 
like DUNE and future very long-baseline proposals can better distinguish 
between environmental asymmetries caused by the Earth and 
fundamental asymmetries caused by CP-violating physics. Several 
concrete directions follow naturally from this work:

\begin{itemize}

    \item \textbf{Full PREM integration.} The four-shell Earth model 
    adopted here is a significant improvement over constant density but 
    remains a simplification. Future work should incorporate the 
    continuous PREM density profile, including the transition zones 
    between shells, to reduce residual modelling bias.

    \item \textbf{Generalisation to inverted ordering:} All results 
    presented here assume normal mass ordering. Since the MSW matter 
    potential affects neutrinos and antineutrinos differently, the 
    baseline-dependent bias is expected to differ under inverted 
    ordering, and a systematic comparison of both hierarchies is 
    warranted.

    \item \textbf{Marginalisation over Earth model uncertainties:} 
    Rather than treating the PREM profile as perfectly known, future 
    analyses should propagate uncertainties in the Earth density 
    profile. They arise from geophysical measurement errors and 
    regional deviations from the spherically symmetric PREM model, as a correlated systematic uncertainty in the $\Delta\chi^2$ fit .

    \item \textbf{Extension to atmospheric neutrinos:} The 
    baseline-dependent bias identified here is directly relevant to 
    atmospheric neutrino analyses with detectors such as 
    IceCube-Upgrade and KM3NeT/ORCA, where neutrinos traverse a 
    continuous distribution of baselines and zenith angles, and 
    where accurate Earth density modelling is essential for 
    extracting $\delta_{CP}$ and the mass ordering simultaneously.

    \item \textbf{Detector resolution effects:} The present analysis 
    assumes perfect energy resolution. Incorporating realistic 
    detector energy smearing would broaden the oscillation features 
    in the event rate spectra, potentially reducing the statistical 
    distinguishability of the two density models and modifying the 
    baseline at which the constant density approximation becomes 
    inadequate.
\end{itemize}

Presently, no proposed long-baseline neutrino experiment has a baseline exceeding approximately 3000 km. Consequently, although the PREM profile may become important for atmospheric neutrinos and very long-baseline studies, its impact on current-generation accelerator-based long-baseline experiments is expected to be comparatively limited.
Ultimately, this integrated approach ensures that future discoveries in the neutrino sector are grounded in a precise understanding of the 
medium through which these particles travel, and that the measurement of $\delta_{CP}$ at next-generation long-baseline experiments is not 
limited by avoidable systematic biases in the Earth density model.


\end{document}